\documentclass{aastex}

\usepackage{graphics,emulateapj5}
\newcommand{\mtot}{M_{\mr{tot}}}
\newcommand{\TT}{\tilde{T}}

\newcommand{\mr}[1]{\mathrm{#1}}

\newcommand{\chandra}{\emph{Chandra}}

\newcommand{\dd}[2]{\frac{d {#1}}{d {#2}}}
\newcommand{\dlndln}[2]{\frac{d \, {\ln{#1}}}{d \, {\ln{#2}}}}

\newcommand{\grad}{\mb{\nabla}}

\newcommand{\mb}[1]{\mathbf{#1}}

\newcommand{\m}{$^{-1}$}

\newcommand{\rhodm}{\rho_\mr{dm}}
\newcommand{\rosat}{\emph{ROSAT}}

\newcommand{\xmm}{\emph{XMM}}

\bibpunct{(}{)}{;}{a}{}{,}%

\begin{document}

\submitted{Submitted December 11, 2002 to the 
\emph{Astrophysical Journal Letters}}

\title{A Gravitational Potential with Adjustable Slope:
A Hydrostatic Alternative to Cluster Cooling Flows}

\author{Andisheh Mahdavi\altaffilmark{1}}
\altaffiltext{1}{amahdavi@ifa.hawaii.edu}

\affil{Institute for Astronomy, University of Hawai'i}

\begin{abstract} 
I discuss a new gravitational potential, $\Phi(r) \propto
(r_0^n+r^n)^{-1/n}$, for modeling the mass distribution of spherical
systems.  This potential has a finite mass and generates a density
profile with inner slope $2-n$. A gas embedded in this potential has
hydrostatic temperature and gas density distributions that are
elementary functions of $n$, greatly simplifying the task of measuring
the slope from X-ray data.  I show that this model is successful in
describing the rising temperature profile and steep gas density
profile often seen in cooling flow clusters. An application to the
Abell 478 cluster of galaxies yields an inner slope $2-n = 1.0 \pm
0.2$ (90\%), consistent with the inner regions of collisionless dark
matter halos first simulated by Navarro, Frenk, and White. The
potential is also useful for cluster dynamics: it is a generalization
of the familiar Hernquist and Plummer potentials, and because it is
invertible, it allows for easy analytic calculation of particle phase
space distribution functions in terms of $n$.
\end{abstract}

\section{Introduction}

The X-ray emitting medium in clusters of galaxies is gravitationally
heated, and therefore a tracer of dark matter. The current generation
of X-ray satellites finally provides the spatial and spectral
resolution necessary to constrain not only the total amount but also
the distribution of matter in clusters. The shape of the mass
distribution is important, because it can be usefully compared with
N-body simulations of structure formation, allowing constraints on
dark matter physics. For example, the inner slope of the dark matter
density, 
\begin{equation}
\nu \equiv -\lim_{r \rightarrow 0} \dlndln{\rhodm}{r}
\end{equation}
can distinguish between collisionless ($\nu \gtrsim 1$) and
self-interacting ($\nu \approx 0$) varieties of dark matter
\citep{Nakano99,Moore99, Burkert00}.

Direct measurements of $\nu$ from X-ray data are rare, however, for
two reasons. The first is that most mass models in common use---e.g.,
the \cite*{NFW} profile $\rho \propto (r_0+r)^2/r$ or the nonsingular
isothermal sphere---do not have an adjustable inner slope $\nu$, and
even if they do, the gas and temperature distributions they generate
are not expressible in terms of elementary functions of $r$ and
$\nu$ \citep{Markevitch97}. This inconvenience makes fitting for $\nu$
an expensive and difficult task.

More importantly, our understanding gas physics in the cores of the
galaxy clusters closest to equilibrium leaves something to be desired.
In the \rosat\ era, such clusters were observed to have cool cores and
cooling times much shorter than the Hubble time; for these reasons
they were thought to possess a quasihydrostatic cooling flow
\citep{Fabian94,Peres98}. However, the \chandra\ and \xmm\ satellites
have in general not detected the multiphase plasma required by cooling
flow models \citep{BohringerM87,Johnstone02,Lewis02}. It is therefore
important at least to ask whether true hydrostatic models are capable
of describing the properties of the gas---specifically the rising
temperature profile---outside the central $\approx 10$ kpc, where the
gas is often disrupted by a central radio source \citep{Blanton01}.

In this Letter I consider a unified solution to the above two
problems. First, I discuss a mass distribution with adjustable inner
slope which does yield hydrostatic gas and temperature profiles in
terms of elementary functions. Then, I show that together with an
exponential pressure distribution, these hydrostatic models can
reproduce the rising temperature profile and the steep density profile
in the rich cluster Abell 478.

\section{ The Potential }

Consider a potential of the form
\begin{equation}
\label{eq:pot}
\Phi(r) = - \frac{G \mtot}{(r_0^n + r^n)^{1/n}}
\end{equation}
Where $\mtot$ is the total mass of the system. The cumulative mass
distribution and density profile that generates this potential are
\begin{equation}
M(r) = \mtot \left( \frac{ r^n } {r^n + r_0^n} \right)^{(1+n)/n}
\label{eq:mass}
\end{equation}
\begin{equation}
\rho(r) = \frac{\mtot (1+n)}{4 \pi r_0^{3}} \frac{(r/r_0)^{n-2}} 
{(r^n/r_0^n + 1)^{2+1/n}}
\end{equation}
The requirement that $M(0) = 0$ yields the lower bound $n > 0$, while
the requirement that the density never increase with radius leads to
an upper bound $n \le 2$. This mass profile therefore corresponds to a
broken power law with inner slope $\nu = 2-n$ and outer slope
$3+n$. The $n = 1$ case corresponds to the \cite{Hernquist90} model,
$n = 2$ is a Plummer sphere, and $n = 1/2$ yields a profile similar to
that of collisionless dark matter halos in recent N-body simulations
\citep{Ghigna00}.

The potential is invertible, 
and thus the density can be expressed as a function of the
potential. If we define a dimensionless potential variable $\psi = -
\Phi r_0 / G M$ then
\begin{equation}
\rho(\psi) = \frac{ \mtot (n+1)}{4 \pi r_0^3} 
\frac{ \psi^{n+3}} {(1 - \psi^n)^{2/n-1}}
\end{equation}

This form of the density can be used to compute phase space
distribution functions using the methods outlined by, e.g.,
\cite{Cuddeford}, and thus facilitate measurement of the mass
profile through equilibrium particle dynamics.

\section{X-Ray Atmospheres of Dark Matter Halos}

Here I investigate the luminosity and temperature distribution of a
hot, X-ray emitting plasma embedded in the mass distribution given by
(\ref{eq:mass}) and governed by the equation of hydrostatic equilibrium:
\begin{equation}
\grad P = - \rho_g \grad \Phi,
\end{equation}
where $P$ is the gas pressure and $\rho_g$ is the gas density, which
in general is different from the total matter density traced by
$\Phi$. With the assumption that the plasma is spherically symmetric,
the equation becomes
\begin{equation}
\frac{1}{\rho_g} 
\label{eq:bhydro}
\dd{\TT \rho_g}{r}  =  - \dd{\Phi}{r} = -\frac{G M}{r^2} 
\end{equation}
where $\TT \equiv P / \rho_g$ is the thermal energy per unit mass. For
an ideal gas in a cluster environment, $\TT = k T / \mu m_p$, where
$T$ is the temperature, $m_p$ is the proton mass, $\mu = 0.61$ is the
mean molecular weight, and $k$ is Boltzmann's constant.

\subsection{Polytropes and the $\beta$-model}

Consider the possibility that the gas is a polytrope, $P \propto \rho^\gamma$, 
where $\gamma$ is the polytropic index.
In this case, the potential yields 
\begin{eqnarray}
\label{eq:tofr}
\label{eq:tbeta}
\TT & = &  \frac{\TT_0}{(r^n/r_0^n + 1)^{1/n}}  \\
\label{eq:beta}
\rho_g & = & \frac{\rho_0}{(r^n/r_0^n + 1)^{3 \beta/n}} \\
\beta & \equiv & \frac{1}{3(\gamma-1)} = \frac{G \mtot}{r_0 \TT_0} -1.
\end{eqnarray}

An important property of this model is that the $n = 2$ potential
reproduces the familiar $\beta$-model, the gas density profile most
widely used by X-ray astronomers.  While most literature assumes that
the $\beta$-model describes an isothermal gas, in the present context
it corresponds to the temperature profile in equation (\ref{eq:tofr}).

Furthermore, the model suggests an important connection between the
polytopic index $\gamma$ and the $\beta$ parameter. It is clear from
equation (\ref{eq:beta}) that, for the gas mass to be finite, we must
have $\beta > 2/3$. On the other hand, for $\beta < 2/3$, the
polytropic index $\gamma > 5/3$, and the plasma becomes subject to
convective instabilities \citep{Nevalainen99}. Thus, two important
physical criteria for the validity of the polytopic model
above---convergence of the gas mass and lack of convective
instabilities---are fulfilled by the very same constraints on $\beta$.

\subsection{A Hydrostatic Alternative to Cooling Flows}

The polytropic models have two marked disadvantages: they require that
the X-ray emitting medium have the same characteristic radius, $r_0$,
as the total matter distribution, which need not in general be true;
they also do not reproduce the temperature structure of the clusters
closest to equilibrium. In many of the most regular clusters of
galaxies, the temperature is observed to rise with distance from the
center, leveling out or falling after a peak. However, it is clear
from equation (\ref{eq:tbeta}) that the temperature in the polytropic
model decreases monotonically with radius, and is therefore unable to
reproduce the rise in temperature observed in many clusters.

For this reason, I introduce a new hydrostatic pressure model that
reproduces many of the properties observed in clusters with rising
temperature profiles. Consider an exponential gas pressure profile:
\begin{equation}
P = P_0 e^{-(r/r_0)^a/b}
\end{equation}
where $P_0$ is the central pressure, and $b^{1/a} r_0$ is the scale
height of the pressure distribution. Together with the equation of
hydrostatic equilibrium and equation (\ref{eq:pot}), this yields the following
gas density and temperature profiles:
\begin{eqnarray}
\rho_g & = & \frac{P_0 a r_0 }{G \mtot b} \left(\frac{r}{r_0} \right)^{a-n}
  \left( 1+ \frac{r^n}{r_0^n} \right)^{1+1/n} e^{-(r/r_0)^a/b} \\
\TT & = &\frac{G \mtot b}{a r_0} \left( \frac{r}{r_0} \right)^{n - a} 
\left( 1+\frac{r^n}{r_0^n} \right)^{-1-1/n}
\end{eqnarray}
In order for the gas density to decrease with the radius, we must
have $n>a$. In this case, the temperature profile behaves
as an increasing power law of index $n-a$, achieving a maximum
at $
r_\mr{max}  =  r_0 [ (n-a)/(1+a) ]^{1/n}$
before falling to zero at infinity. When $n=a$, the temperature peak
is at $r=0$, giving us again a monotonically decreasing temperature
profile.

\begin{figure*}
\resizebox{7in}{!}{\includegraphics{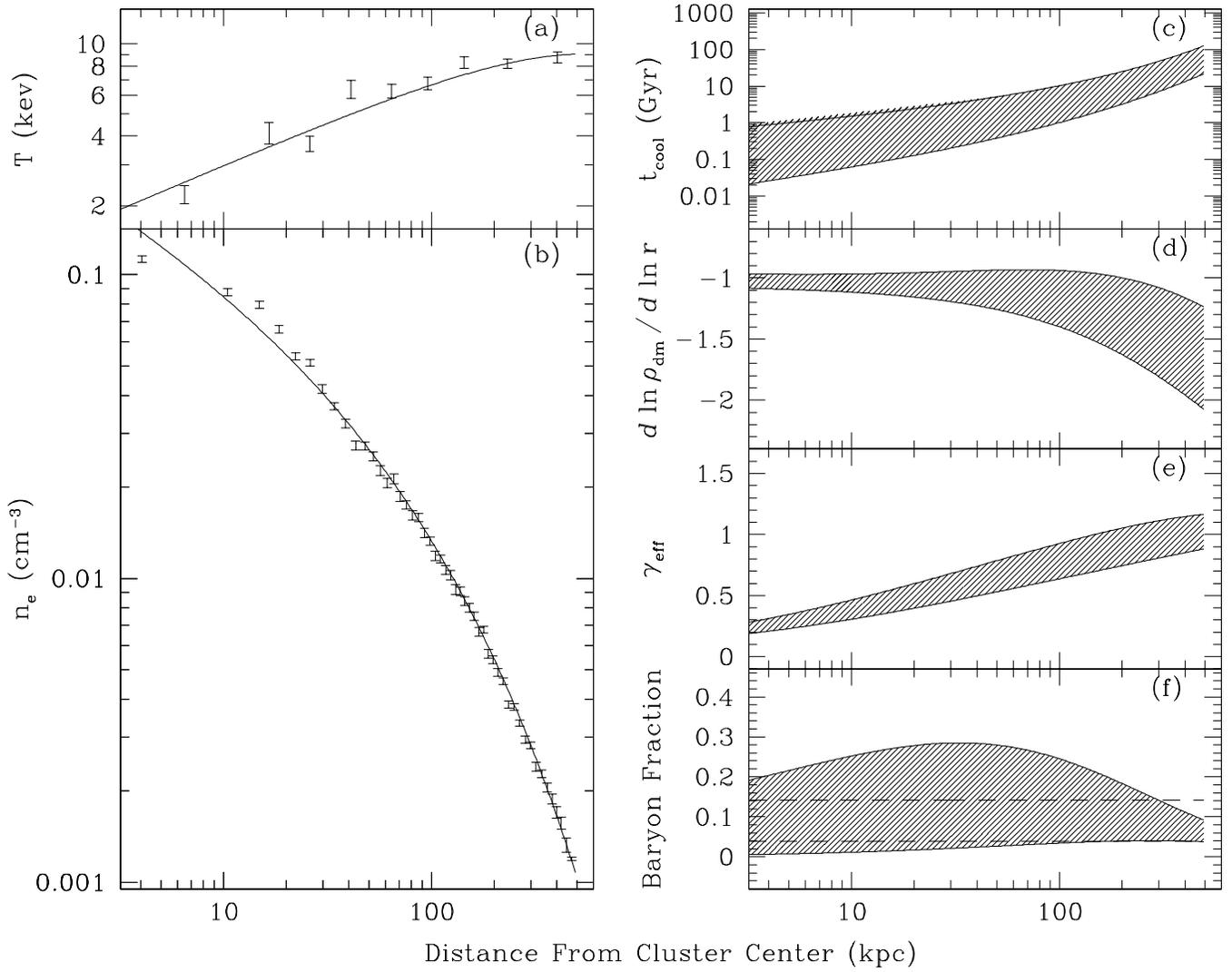}} \figcaption{Fits of the
present model to the temperature and electron density distribution of
Abell 478. The data are taken from \cite{Sun02}; the model fits are
mine. (a) Deprojected temperature profile; (b) deprojected
electron density. Shown also are 68\% confidence limits on the
following quantities as a function of radius: (c) the cooling time;
(d) the slope of the dark matter density $\nu = d \ln{\rhodm} / d
\ln{r}$; (e) the effective polytropic index $d \ln{P} / d \ln
{\rho_g}$; and (f) the baryon fraction (solid lines)
as well as on the average baryon fraction with 0.5 Mpc (dashed
horizontal lines).
\label{fig:a478}}
\end{figure*}

In Figure \ref{fig:a478}, I show that this model fits the temperature
and the emissivity profile of Abell 478 as observed by the \chandra\
satellite.\footnote{Assuming $H_0 = 70$ km s\m\ Mpc\m, $\Omega_M =
0.3$, and $\Omega_\Lambda = 0.7$} The deprojected data are from
\cite{Sun02}; all analysis and model fitting is mine. The minimized
six-dimensional merit function is
\begin{equation}
\chi^2 = \sum_i^{N_n} \left[ \frac{n_i - \rho_g(r_i)/m_p}{\Delta n_i} \right]^2 +
         \sum_i^{N_T} \left[ \frac{T_i - \TT(r_i)}{\Delta T_i} \right]^2,
\end{equation}
where $N_n$ is the total number of density data points $n_i$, and
$N_T$ is the total number of temperature data points $T_i$. The
minimized value is 39.0 for 49 degrees of freedom, indicating that
the fit is of a high quality.

\begin{figure*}
\resizebox{7in}{!}{\includegraphics{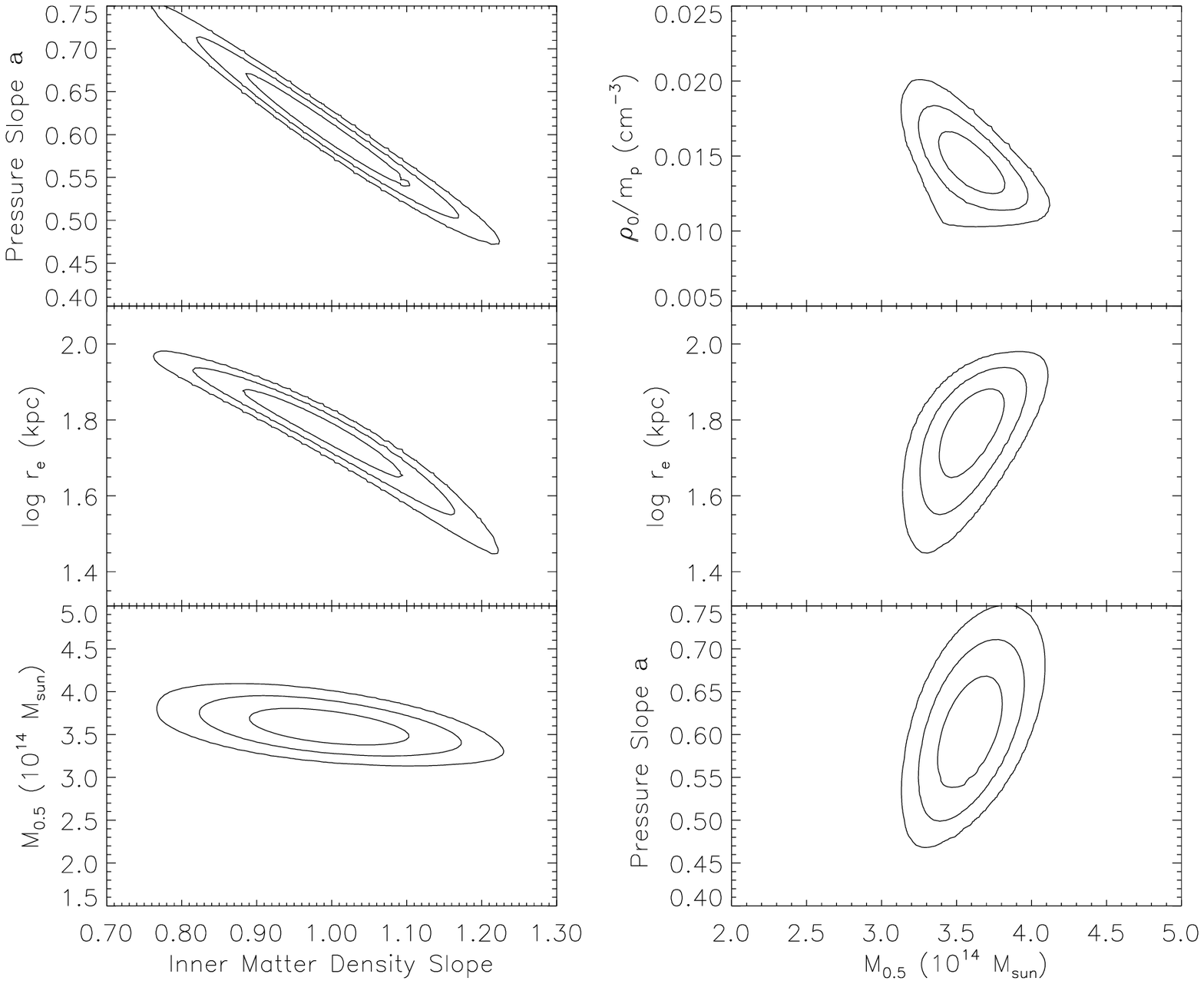}} \figcaption{Marginalized
Bayesian probability distributions for 6 of the 15 unique parameter
pairs. Shown are the 68\%, 95\%, and 99.7\% confidence contours.
\label{fig:conf}}
\end{figure*}

There are 6 total parameters, and therefore $6 \times 5 /2 = 15$
unique pairs of parameters. I derive marginalized joint probability
distributions for all these pairs via a four-dimensional integration
of the Bayesian probability distribution $\exp{(-\chi^2/2)}$, with
uniform prior. Six of these marginalized distributions are shown in
Figure \ref{fig:conf}. To obtain closed confidence contours, it was
necessary to substitute, without loss of generality, the parameters
$r_e \equiv b^{1/a} r_0$ and $M_{0.5} \equiv M(<0.5\mr{\ Mpc}). $
These substitutions are sensible, as $r_e$ is just the exponential
scale height of the gas, and $M_{500}$ is the mass within a spherical
radius of 500 kpc---about the extent of the \chandra\ field of view.

Finally, in order for the previous discussion to be physically
self-consistent, three criteria must apply. First, the gas density
must never exceed the total matter density. This corresponds to the
requirement that the baryon fraction $\rho_g / \rho < 1$ everywhere
throughout the cluster. Second, the gas must not experience convective
instabilities (-$d \ln{\rho_g} / d \ln{r} \lesssim 5/3$). Third, the
inner slope of the dark matter density $\rhodm = \rho - \rho_g$ must
be checked against the inner slope of the total density $\rho$. Figure
\ref{fig:a478} shows that all these criteria are abundantly
fulfilled. The average baryon fraction within 0.5 Mpc is $0.08 \pm
0.07$ (at 68\% confidence), and the radial variation of it within that
range is equally well-constrained. The effective polytropic index is
always $< 5/3$, and the slope of the dark matter density profile
matches that of the total profile. Hence the model as applied to Abell
478 is self-consistent.

Of course, any hydrostatic model has to contend with the fact that the
cooling time remains quite short. Figure \ref{fig:a478} shows that the
cooling time on the average is smaller than the Hubble time in the
region observed by \chandra, in agreement with \cite{Sun02}; the gas
should radiate away all its energy in under ten billion
years. However, the failure of \chandra\ and \xmm\ observations to
detect the temperature distribution predicted by cooling flow models
\citep{Lewis02,Peterson01} has lead some to infer that an unknown
heating mechanism must balance the cooling
\citep{Loeb02,Bruggen02}. While the details of this balancing system
are beyond the scope of this Letter, its net effect may be to make the
hydrostatic solutions empirically valid in at least some cases.

\vfill
\section{Conclusion}

I suggest a new potential, $\Phi \propto (r^n+r_0^n)^{-1/n}$, for
modeling the matter distribution of galaxy clusters. This potential
has an inner density cusp of slope $2-n$, and is particularly useful
because (1) it is a generalization of potentials in common use; (2) it
is invertible; and (3) a gas embedded in it can have density and
temperature distributions that are elementary functions of $n$. As an
example, I successfully apply a hydrostatic model to published
\chandra\ observations of the galaxy cluster Abell 478, which is
generally believed to contain a cooling flow. The resulting inner slope
$2-n = 1.0 \pm 0.2$ (90\% confidence interval) is well constrained,
and consistent with N-body simulations in which massive halos are
composed of collisionless matter. 

This work demonstrates that cold cores and steep density gradients are
compatible with hydrostatic models in all respects except for the
cooling time.

I am grateful to Margaret Geller, Pat Henry, and Kathleen Kang for
valuable comments on this Letter, and to to Ming Sun for providing me
with the deprojected \chandra\ data. This research was supported by
NASA through a Chandra Postdoctoral Fellowship Award issued by the
Chandra X-ray Observatory Center, which is operated by the Smithsonian
Astrophysical Observatory for and on behalf of NASA under contract NAS
8-39073.

\end{document}